\documentclass[11pt]{article}

\usepackage[letterpaper,margin=1in]{geometry}
\usepackage[T1]{fontenc}
\usepackage[utf8]{inputenc}
\usepackage{mathptmx}
\usepackage[scaled=0.92]{helvet}
\usepackage{courier}
\usepackage{microtype}
\usepackage{amsmath,amssymb}
\usepackage{booktabs}
\usepackage{tabularx}
\usepackage{multirow}
\usepackage{float}
\usepackage{graphicx}
\usepackage{xcolor}
\usepackage{tikz}
\usetikzlibrary{arrows.meta,positioning,fit,backgrounds,calc}
\usepackage{pgfplots}
\pgfplotsset{compat=1.18}
\usepgfplotslibrary{groupplots}
\usepackage[numbers,sort&compress]{natbib}
\usepackage{url}
\usepackage[hidelinks]{hyperref}
\usepackage[capitalise,noabbrev]{cleveref}
\usepackage{caption}
\definecolor{seriesA}{HTML}{2A78D6}
\definecolor{seriesB}{HTML}{EB6834}
\definecolor{seriesC}{HTML}{1BAF7A}
\definecolor{inkmuted}{HTML}{6B6B66}
\definecolor{gridgray}{HTML}{E4E3DE}

\pgfplotsset{
  paperaxis/.style={
    axis lines=left,
    tick style={draw=none},
    major grid style={draw=gridgray},
    tick label style={font=\footnotesize, text=black},
    label style={font=\footnotesize},
    legend style={font=\footnotesize, draw=none, fill=none},
    every axis plot/.append style={line width=1pt},
  }
}

\hypersetup{
  pdftitle={Characterizing AI Service Traffic for 6G: An Open Testbed and Initial Results from 3GPP SA4},
  pdfauthor={Imed Bouazizi},
  pdfsubject={AI service traffic characterization and network emulation},
  pdfkeywords={6G, AI traffic, network emulation, 3GPP SA4}
}

\title{Characterizing AI Service Traffic for 6G:\\
An Open Testbed and Initial Results from 3GPP SA4}

\author{Imed Bouazizi\\
Qualcomm Technologies, Inc.\\
\texttt{bouazizi@qti.qualcomm.com}}

\date{October 2026}

\begin{document}
\maketitle

\begin{abstract}
The design of sixth generation (6G) mobile systems requires measurements of artificial intelligence (AI) service traffic under controlled network conditions. The 3rd Generation Partnership Project (3GPP) Service and System Aspects Working Group 4 (SA4) documents such measurements in draft Technical Report 26.870. In this paper, we describe the open testbed used for the study and analyze the results recorded in version 0.6.2 and its companion workbook. The testbed combines network emulation, packet capture, application measurements, and scenario execution against hosted and self-hosted models. The reported uplink to downlink volume ratios span more than six orders of magnitude. A personal assistant agent has a ratio of 0.007 at its local control interface and 0.50 on its wide-area egress. Deterministic agent signaling follows the configured round-trip delay. In a tokenized video experiment with 125 questions per profile, the number of correct answers changes by at most three through 10\% independent uplink loss. Larger losses reduce accuracy, but the available repetitions do not establish a tolerance threshold. We also analyze emulator behavior that affects the interpretation of the study. Per-packet jitter introduces reordering, correlated random-loss settings suppress rare loss events, and a Gilbert-Elliott parameter mapping changes the stationary loss rate. The results support characterization by scenario, measurement layer, and versioned network configuration.
\end{abstract}
\section{Introduction}
\label{sec:intro}

The 3rd Generation Partnership Project (3GPP) is studying artificial intelligence (AI) services as part of its sixth generation (6G) work. Technical Report (TR) 22.870~\cite{tr22870} identifies use cases that include conversational assistants, agents, and media inference. Radio Access Network Working Group 2 (RAN2) has asked Service and System Aspects Working Group 4 (SA4) to characterize their directionality, burstiness, delay sensitivity, variability, and token transport. These questions require measurements that participating companies can reproduce. In this paper, we describe the open testbed used for the SA4 study and analyze the results recorded in draft TR 26.870 V0.6.2~\cite{tr26870}.

Commercial services expose only part of the system. Their traffic is encrypted, their inference configuration may change, and the measurement client does not control the external network. A trace from one deployment cannot isolate these effects. The testbed runs a declared interaction pattern against a declared model and network profile. Application records and packet captures describe the same experiment. Hosted execution measures the service as deployed. Local execution provides more control over the inference engine and the emulated path.

The contributions of this paper are the following:
\begin{itemize}
  \item We describe a testbed architecture that separates network emulation and packet analysis from AI scenario execution. A dataset component consumes the resulting captures and session records.
  \item We describe the measurement methodology used in the SA4 study, including scenarios, network profiles, execution modes, and the association of application records with packet evidence.
  \item We analyze the results recorded in the draft report and workbook for the five RAN2 question families. We distinguish reported measurements from derived ratios and their interpretation.
  \item We analyze emulator behavior that changes the conditions produced by a profile. We identify the resulting limits on comparisons across campaigns.
\end{itemize}

The remainder of the paper is organized as follows. \Cref{sec:related} reviews related work. \Cref{sec:design} describes the design principles and the architecture of the testbed. \Cref{sec:method} describes scenarios, network profiles and metrics. \Cref{sec:results} summarizes the results of TR 26.870. \Cref{sec:fidelity} discusses emulator fidelity and other threats to validity. \Cref{sec:standards} describes the use of the testbed in standardization, and \cref{sec:conclusion} summarizes the results and further work.
\section{Related Work}
\label{sec:related}

\paragraph{Traffic models in 3GPP.}
3GPP has developed traffic models for extended reality (XR). TR 38.838~\cite{tr38838} defines frame-size, periodicity, and jitter models for XR evaluations in New Radio (NR). The 5G system architecture in TS 23.501~\cite{ts23501} includes protocol data unit (PDU) set handling for XR services. AI applications add interaction patterns governed by model execution, tool calls, and context transfer. The measurement methodology described here provides data for evaluating these patterns.

\paragraph{Workload characterization of LLM serving.}
Several studies characterize large language model (LLM) and agentic workloads from the serving side. Yuan et al.~\cite{yuan2026agentic} show that effective context caching allows agents to reuse input tokens across turns. Chang et al.~\cite{chang2026agentic} show that tools, execution environments, and state management dominate latency in several evaluated applications. These studies explain serving costs and application execution. Access-path measurements address an additional part of the interaction.

\paragraph{Network-level traffic of generative AI.}
Nguyen et al.~\cite{nguyen2026mint} collect client-side traces from three LLM providers across four modalities and develop burst models for ns-3 simulation. Lamagna et al.~\cite{lamagna2026mas} show that agent coordination topology changes the arrival process of LLM requests. Their study includes inference and network measurements. Kundu et al.~\cite{kundu2026ainative} use the 5G-MAG testbed to characterize AI traffic for 6G awareness mechanisms. Our work describes the testbed itself, its use in the SA4 methodology, and the interpretation of measurements collected under declared impairment profiles.

\paragraph{Traffic analysis and privacy.}
Encrypted AI traffic can expose information through packet size, direction, and timing. Weiss et al.~\cite{weiss2024keylogging} study response reconstruction from assistant traffic. Pouryousef et al.~\cite{pouryousef2026sidechannel} evaluate model and prompt-category identification from encrypted metadata. Li et al.~\cite{li2026ant} provide a traffic dataset for agent behavior auditing. These studies show that packet features retain application structure. Replacing provider and model names with neutral labels does not remove the information that traffic analysis may recover. Sharing captures requires a separate review of payloads, addresses, identifiers, and metadata.

\paragraph{Network emulation.}
NetEm~\cite{hemminger2005netem} is the de facto standard for emulating delay, jitter, loss and rate on Linux. Its accuracy and limitations have been studied by Nussbaum and Richard~\cite{nussbaum2009emulators} and by Jurgelionis et al.~\cite{jurgelionis2011netem}. We revisit two of its behaviors, reordering under per-packet jitter and the semantics of correlated loss, because they directly affect the profiles used in the 3GPP study.
\section{Testbed Design}
\label{sec:design}

\subsection{Design Principles}

The testbed was built to serve a standards study rather than a single experiment. That purpose leads to the following design principles:
\begin{itemize}
  \item \textbf{Conditional reporting.} A metric is associated with a scenario and a network profile. An aggregate across profiles states its population and weighting.
  \item \textbf{Reproducibility.} Configuration files identify scenarios, models, and profiles. Session identifiers associate the configuration with the application record and packet capture.
  \item \textbf{Measurement layers.} Application records and packet captures describe the same session. Their byte counts retain the accounting rules of their respective layers.
  \item \textbf{Separation of layers.} Network shaping, capture and packet analysis are independent of the AI experiments, so that other studies can reuse the network layer without the AI scenarios.
  \item \textbf{Result sharing.} Provider and model identifiers can be replaced with neutral labels in exported results. Capture sharing requires additional data review.
\end{itemize}

\subsection{Architecture}

The testbed consists of three components. The \texttt{netemu} package provides network emulation, capture, and analysis of packet capture (PCAP) files. The \texttt{aitestbed} component drives experiments and depends on \texttt{netemu}. The \texttt{training} component reads capture files and session records without importing the experiment code. The following figure depicts the architecture.

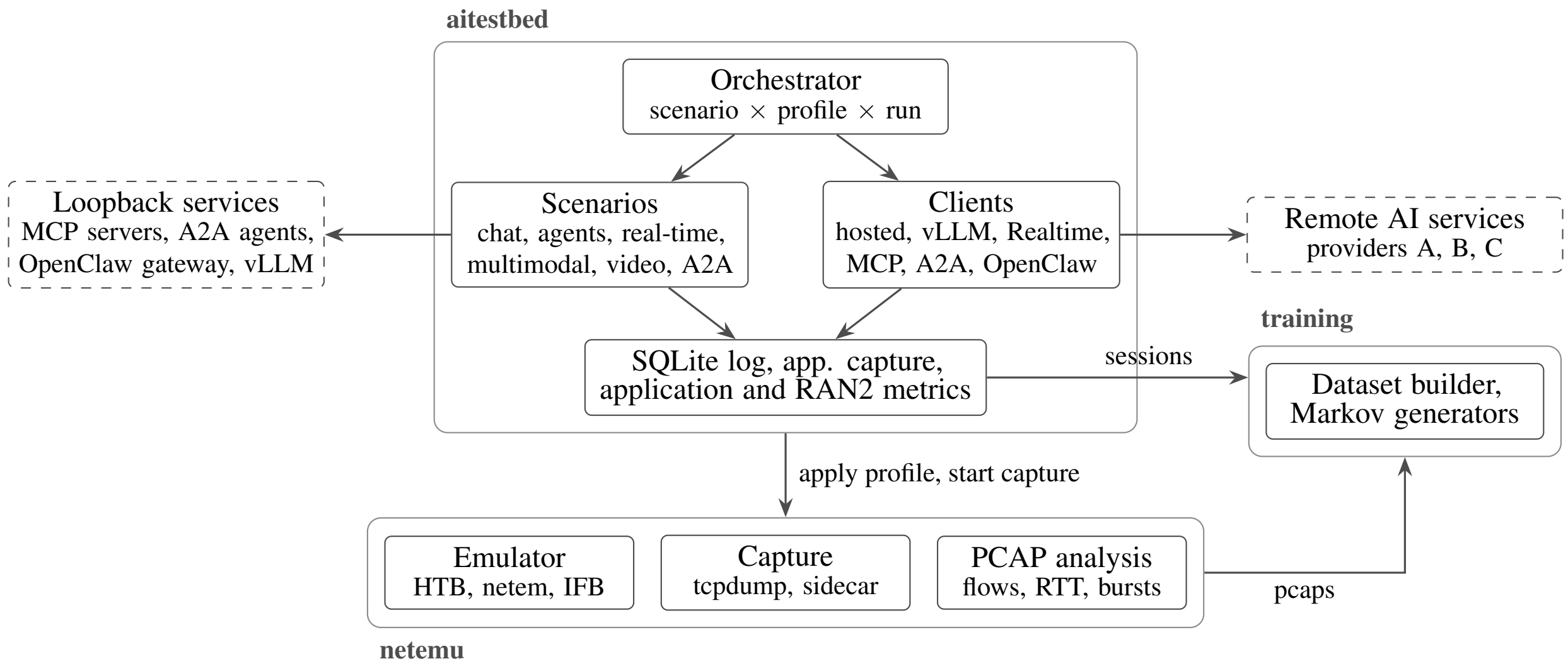
\begin{figure}[!htbp]
\centering
\resizebox{\linewidth}{!}{
\begin{tikzpicture}[
  font=\footnotesize,
  box/.style={font=\scriptsize, draw=black!70, rounded corners=2pt, align=center, fill=white, inner sep=3pt, text width=#1},
  box/.default=26mm,
  grp/.style={draw=black!45, rounded corners=4pt, inner sep=5pt},
  ext/.style={font=\scriptsize, draw=black!70, dashed, rounded corners=2pt, align=center, fill=white, inner sep=3pt, text width=31mm},
  arr/.style={-{Stealth[length=2mm]}, draw=black!70, line width=0.6pt},
  lbl/.style={font=\scriptsize\bfseries, text=black!70},
  sub/.style={font=\scriptsize}
]
\node[box=32mm] (orch) at (7.5,0) {{\footnotesize Orchestrator}\\[-1pt]{scenario $\times$ profile $\times$ run}};
\node[box=29mm] (scen) at (5.55,-1.45) {{\footnotesize Scenarios}\\[-1pt]chat, agents, real-time,\\multimodal, video, A2A};
\node[box=29mm] (cli) at (9.45,-1.45) {{\footnotesize Clients}\\[-1pt]hosted, vLLM, Realtime,\\MCP, A2A, OpenClaw};
\node[box=40mm] (log) at (7.5,-2.95) {\footnotesize SQLite log, app. capture,\\[-1pt]\footnotesize application and RAN2 metrics};
\begin{scope}[on background layer]
\node[grp, fit=(orch)(scen)(cli)(log)] (gA) {};
\end{scope}
\node[lbl, anchor=south west] at (gA.north west) {aitestbed};
\node[ext] (loc) at (1.0,-1.45) {{\footnotesize Loopback services}\\[-1pt]MCP servers, A2A agents,\\OpenClaw gateway, vLLM};
\node[ext] (rem) at (14.0,-1.45) {{\footnotesize Remote AI services}\\[-1pt]providers A, B, C};
\node[box=24mm] (emu) at (4.6,-5.0) {{\footnotesize Emulator}\\[-1pt]{HTB, netem, IFB}};
\node[box=24mm] (cap) at (7.5,-5.0) {{\footnotesize Capture}\\[-1pt]{tcpdump, sidecar}};
\node[box=24mm] (pcap) at (10.4,-5.0) {{\footnotesize PCAP analysis}\\[-1pt]{flows, RTT, bursts}};
\begin{scope}[on background layer]
\node[grp, fit=(emu)(cap)(pcap)] (gN) {};
\end{scope}
\node[lbl, anchor=north west] at (gN.south west) {netemu};
\node[box=27mm] (train) at (14.0,-3.2) {\footnotesize Dataset builder,\\[-1pt]\footnotesize Markov generators};
\begin{scope}[on background layer]
\node[grp, fit=(train)] (gT) {};
\end{scope}
\node[lbl, anchor=south west] at (gT.north west) {training};

\draw[arr] (orch) -- (scen);
\draw[arr] (orch) -- (cli);
\draw[arr] (scen) -- (log);
\draw[arr] (cli) -- (log);
\draw[arr] (scen.west) -- (loc.east);
\draw[arr] (cli.east) -- (rem.west);
\draw[arr] (gA.south) -- node[right, sub]{apply profile, start capture} (gN.north);
\draw[arr] (log.east) -- node[above, sub, pos=0.62]{sessions} (gT.west |- log.east);
\draw[arr] (gN.east |- pcap.east) -- node[below, sub, pos=0.5]{pcaps} (pcap.east -| gT.south) -- (gT.south);
\end{tikzpicture}
}
\caption{Architecture of the testbed. Solid boxes are testbed components. Dashed boxes are services that the scenarios reach over the loopback interface or over the wide-area network.}
\label{fig:arch}
\end{figure}

\subsection{Network Emulator}

The network emulator wraps Linux traffic control (tc). On the main interface, a hierarchical token bucket (HTB) class limits rate and a netem queue applies delay, jitter, loss, reordering, duplication, and corruption. An intermediate functional block (IFB) device provides ingress shaping. A profile can override egress parameters for asymmetric links. Loopback uses a separate queue and netem's native rate option. These paths can produce different delay and ordering behavior under the same nominal profile. Local peers include Model Context Protocol (MCP)~\cite{mcp} servers, Agent2Agent (A2A)~\cite{a2a} agents, an agent gateway, and self-hosted inference.

Packet capture runs tcpdump and records a metadata sidecar containing the scenario, profile, and run identifiers. The configured filter covers Hypertext Transfer Protocol (HTTP), HTTP over Transport Layer Security (HTTPS), and selected Web Real-Time Communication (WebRTC) service ports. The latter include ports used for connectivity checks, relays, and media in the evaluated deployments. A service-port filter does not cover every dynamically negotiated media port. Each new deployment needs a capture-coverage check.

The analysis module parses libpcap and pcapng captures containing Internet Protocol version 4 (IPv4) and version 6 (IPv6) traffic. Transmission Control Protocol (TCP) handshakes determine client and server roles when available. This avoids relying on the ordering of port numbers for loopback services. The module derives handshake round-trip time (RTT), Transport Layer Security (TLS) setup duration, retransmission indicators, directional byte volumes, throughput, and burst statistics.

\subsection{Experiment Orchestration}

The orchestrator iterates over a test matrix of scenarios and profiles. For each pair, it applies the profile, starts the captures, executes the configured number of runs and stores one record per request in an SQLite database. A record carries request and response sizes, token counts, the request start time and the arrival times of the first and last token, inter-chunk gaps for streamed responses, tool call statistics, HTTP status and error type.

Scenarios share a common base class and are registered by type. Clients adapt them to model providers. The inspected configuration contains more than 60 scenario definitions. They cover hosted providers, a self-hosted vLLM server~\cite{vllm}, real-time conversation, MCP tools, A2A agents, and the OpenClaw runtime~\cite{openclaw}. An optional application-layer capture through mitmproxy~\cite{mitmproxy} records decrypted HTTP exchanges with configured redaction. Definitions in the catalog are distinct from the subset executed in each reported campaign.

The orchestrator supports resuming an interrupted campaign. A session counts toward completion when all its records succeed or when it is recorded as a timeout. Other failed sessions may be retried. A timeout is retained as an outcome, although its cause may include provider or host behavior as well as the network. Analysis needs to retain attempts and retries separately. A completed-run count alone is not a denominator for service reliability.

\subsection{Synthetic Traffic Generation}

The dataset component associates flows with transport surfaces and session windows. Traffic between loopback endpoints is kept separate from wide-area egress. A persistent connection that spans sessions is segmented at their recorded boundaries. Flows receive one of nine traffic-category labels. Training, validation, and test partitions are assigned by capture group, including paired interface captures. This prevents packets from one capture group from appearing in several partitions. It does not establish independence across campaigns, providers, or repeated prompts.
\section{Measurement Methodology}
\label{sec:method}

\subsection{Scenarios}

A scenario defines an interaction pattern through its input and output modalities, interaction style, protocol, and model configuration. Annex D of draft TR 26.870 maps the evaluated scenarios to use cases in TR 22.870~\cite{tr26870}. Server-sent events (SSE), Real-time Transport Protocol (RTP), and JavaScript Object Notation remote procedure calls (JSON-RPC) identify specific protocol forms used by these scenarios. The following table shows the scenario families.

\begin{table}[H]
\centering
\small
\caption{Scenario families of the testbed and their dominant traffic pattern.}
\label{tab:scenarios}
\begin{tabularx}{\linewidth}{@{}lXl@{}}
\toprule
Family & Interaction & Transport \\
\midrule
Chat & Single and multi-turn chat, streaming and non-streaming, coding, reasoning, token identifiers & HTTPS, SSE \\
Agent with MCP tools & Iterative LLM and tool rounds for shopping, research, market data, music & HTTPS, MCP over HTTP \\
Browser and computer use & Screenshot-driven control and Playwright automation & HTTPS \\
Image generation & Text prompt, generated image & HTTPS \\
Multimodal analysis & Image and text upload, streamed or single response & HTTPS, SSE \\
Video understanding & Base64 video upload, or tokenized video over WebRTC with data channel response & HTTPS, RTP \\
Real-time conversation & Text and voice assistants & WebSocket, WebRTC \\
Direct web search & Multi-threaded search with LLM synthesis & HTTPS \\
Personal assistant agent & Local agent runtime driving a hosted model and tools & Loopback, HTTPS \\
Agent-to-agent & Single task, streaming task, multi-agent delegation & JSON-RPC, SSE \\
\bottomrule
\end{tabularx}
\end{table}

Two execution modes identify where inference runs. In remote execution, the client reaches a hosted model through the main interface. Emulation adds impairment to an external path that the testbed does not control. In local execution, the client reaches a self-hosted model through loopback, where a separate queue applies impairment. Local execution reduces provider-side variation. It still includes inference variability, host scheduling, and the semantics of a queue shared by both directions. It does not emulate a radio protocol stack.

\subsection{Network Profiles}

The profile catalog combines deployment reference points and stress cases. SA4 agreed profile inputs in S4-260848~\cite{s4260848}. The \texttt{6g\_itu\_hrllc} profile uses a reference point motivated by the hyper reliable and low-latency communication (HRLLC) objectives in ITU-R M.2160~\cite{itum2160}. It does not implement or validate an IMT-2030 radio system. The 5G quality of service (QoS) identifier (5QI) anchors use $d = \mathrm{PDB} - 2.054\,\sigma$ for selected normal-delay inputs, where PDB denotes the packet delay budget. The coefficient is the standard-normal 98th percentile. This construction does not guarantee the observed path percentile after shaping and queueing. The following table shows the nominal inputs used to identify the study profiles.

\begin{table}[H]
\centering
\footnotesize
\caption{Nominal profile inputs. Delay and jitter are netem location and scale parameters in each direction. Asymmetric entries show downlink / uplink. Realized conditions depend on the shaping path and workload.}
\label{tab:profiles}
\begin{tabularx}{\linewidth}{@{}lrrXrll@{}}
\toprule
Profile & Delay (ms) & Jitter (ms) & Loss model & Loss (\%) & Rate (Mbit/s) & Role \\
\midrule
\texttt{no\_emulation} & 0 & 0 & none & 0 & -- & Baseline \\
\texttt{6g\_itu\_hrllc} & 1 & 0.2 & correlated & 0.001 & 300 & IMT-2030 HRLLC \\
\texttt{5g\_urban} & 20 & 5 & correlated & 0.1 & 100 & Urban cellular \\
\texttt{wifi\_good} & 30 & 10 & correlated & 0.1 & 50 & Non-3GPP access \\
\texttt{cell\_edge} & 120 & 30 & Gilbert-Elliott & 1 & 5 & Weak coverage \\
\texttt{satellite\_leo} & 22 & 7 / 8 & correlated & 0.5 / 0.8 & 100 / 15 & LEO satellite \\
\texttt{satellite\_geo} & 340 & 15 / 18 & correlated & 0.1 / 0.2 & 50 / 3 & GEO satellite \\
\texttt{congested} & 200 & 50 & Gilbert-Elliott & 3 & 1 & Bufferbloat \\
\texttt{5qi\_7} & 80 & 10 & correlated & 0.1 & -- & 5QI 7 anchor \\
\texttt{5qi\_80} & 8 & 1 & correlated & 0.0001 & -- & 5QI 80 anchor \\
\texttt{lossy} & 0 & 0 & independent & 1 to 30 & -- & Loss tolerance \\
\bottomrule
\end{tabularx}
\end{table}

The \texttt{lossy} profile applies independent loss without added delay. The sweep from 1\% to 30\% evaluates failure behavior rather than expected deployment conditions. Annex D excludes this profile from aggregates intended to describe deployment conditions. The report and workbook retain some other aggregates that include it. Each result below identifies the population it uses.

\subsection{Metrics}

The metrics fall into four categories: traffic characteristics, response time, service metrics and task-related quality of experience (QoE). The time to first token (TTFT) and the time to last token (TTLT) are computed from the per-request record as
\begin{equation}
  \mathrm{TTFT} = t_{\mathrm{first}} - t_{\mathrm{start}}, \qquad
  \mathrm{TTLT} = t_{\mathrm{last}} - t_{\mathrm{start}},
\end{equation}
where $t_{\mathrm{start}}$ is the request issue time and $t_{\mathrm{first}}$ and $t_{\mathrm{last}}$ are the first and last recorded response-chunk arrival times. For a non-streamed response, the two arrival times coincide. A response chunk may contain several model tokens.

Burst structure is computed from the packet capture. Packets of one direction are walked in time order, and a burst is closed when the inter-arrival time exceeds a gap threshold $g$, set to 10~ms or 100~ms. The burstiness index at window $w$ is
\begin{equation}
  \beta_w = \frac{\max_k S_k}{\frac{1}{K}\sum_{k=1}^{K} S_k},
\end{equation}
where $S_k$ is the number of bytes in the $k$-th window of length $w$ and $K$ the number of windows. The index is evaluated for $w$ from 1~ms to 10~s. Windows below 10~ms are affected by scheduling jitter of the host operating system and are treated as indicative.

The methodology to answer RAN2 was agreed in S4-260859~\cite{s4260859}. It organizes the network-layer metrics into five question families, one per assumption: uplink-heavy traffic, bursts and delay bounds, round-trip delay, intra-application variability and tokenized traffic. Several of these metrics join the application-layer records to the packet captures of the same session. For example, the relation between tokens and bytes is estimated per scenario by least squares,
\begin{equation}
  b_{\mathrm{UL}} \approx \alpha\, n_{\mathrm{in}} + \beta, \qquad
  b_{\mathrm{DL}} \approx \gamma\, n_{\mathrm{out}} + \delta,
\end{equation}
where $b$ denotes the payload bytes of a turn and $n$ its token count.

For scenarios whose output is consumed by a machine rather than a person, delivery metrics are not enough. A video understanding service may deliver every packet in time and still answer the question wrongly. The task-related QoE metric for closed-form video understanding is the accuracy
\begin{equation}
  A_{\mathrm{vu}} = \frac{N_{\mathrm{correct}}}{N_{\mathrm{questions}}},
\end{equation}
over a fixed question set. The denominator includes the questions in the evaluation. Format-invalid responses are also recorded as a separate diagnostic. Their treatment needs to remain fixed when accuracy is compared across profiles.
\section{Results from TR 26.870}
\label{sec:results}

This section analyzes measurements recorded in draft TR 26.870 V0.6.2 and the workbook distributed with S4aP260148~\cite{tr26870}. Providers retain the workbook labels A, B, and C. The commercial-service summary contains 14,355 request records across ten profiles, including the unshaped reference. The associated flow summary reports 104,664 TCP flows from 1,320 captures. Local execution covers chat, reasoning, multimodal analysis, and video understanding. The agentic worksheets report 200 personal-assistant turns and 198, 99, and 100 turns for the three A2A variants across ten profiles~\cite{s4261814}. These populations are separate campaigns. The values below are reported measurements or calculations from the workbook, rather than a new measurement campaign.

\subsection{Directionality}

The first RAN2 question concerns uplink-heavy traffic. We derive the directional ratio as mean request bytes divided by mean response bytes. This ratio of means differs from a mean of per-turn ratios. Agent network-layer measurements are reported separately below. The following figure depicts application-layer ratios from the workbook summary.

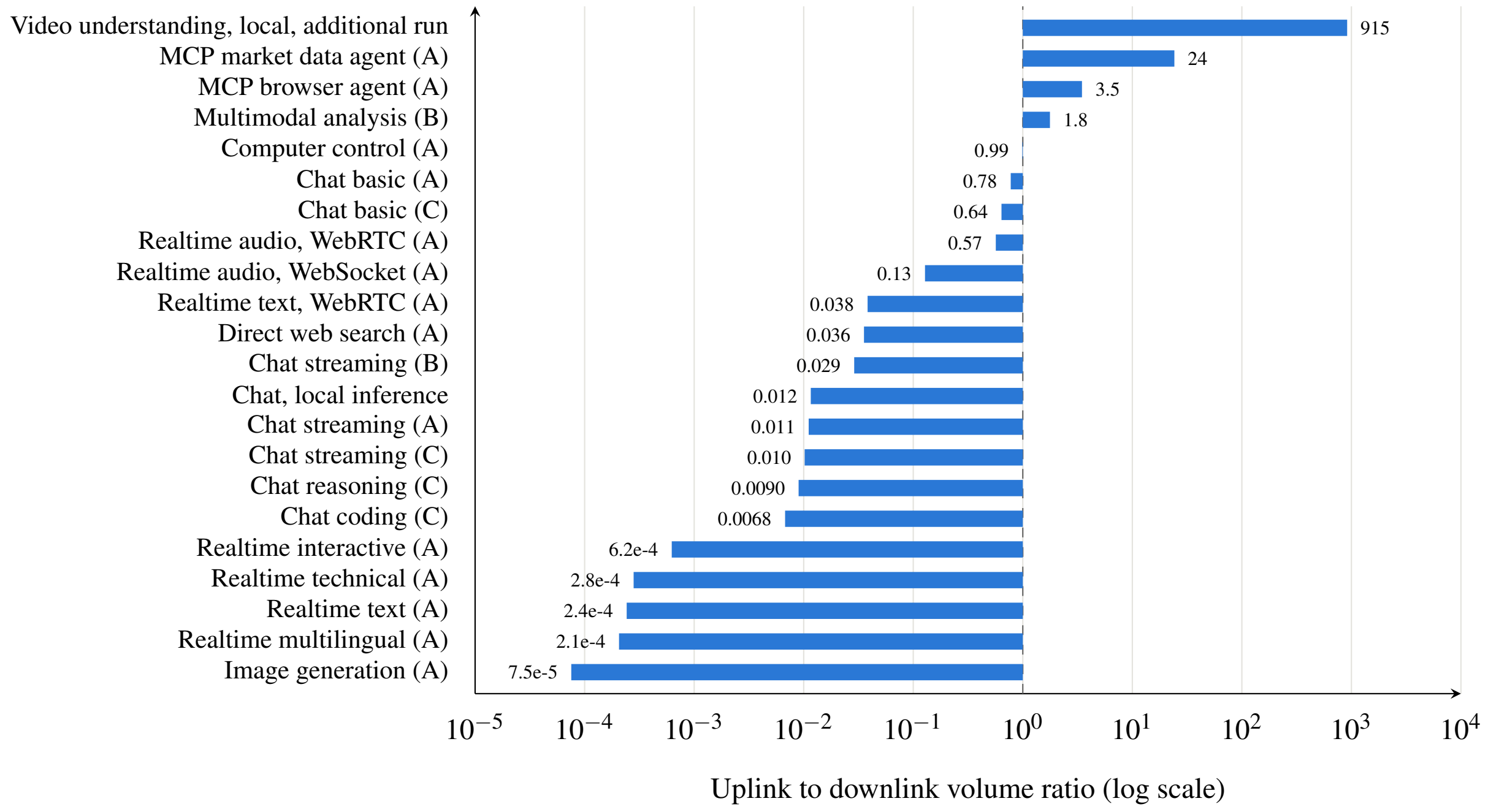
\begin{figure}[!htbp]
\centering
\begin{tikzpicture}
\begin{axis}[
  paperaxis,
  xbar, bar width=5pt,
  width=0.74\linewidth, height=9cm,
  xmin=1e-5,
  xmode=log, log basis x=10,
  xmax=1e4,
  ymin=0.3, ymax=22.7,
  xlabel={Uplink to downlink volume ratio (log scale)},
  ytick={1,...,22},
  yticklabels={
    {Image generation (A)},
    {Realtime multilingual (A)},
    {Realtime text (A)},
    {Realtime technical (A)},
    {Realtime interactive (A)},
    {Chat coding (C)},
    {Chat reasoning (C)},
    {Chat streaming (C)},
    {Chat streaming (A)},
    {Chat, local inference},
    {Chat streaming (B)},
    {Direct web search (A)},
    {Realtime text, WebRTC (A)},
    {Realtime audio, WebSocket (A)},
    {Realtime audio, WebRTC (A)},
    {Chat basic (C)},
    {Chat basic (A)},
    {Computer control (A)},
    {Multimodal analysis (B)},
    {MCP browser agent (A)},
    {MCP market data agent (A)},
    {Video understanding, local, additional run}
  },
  yticklabel style={font=\scriptsize},
  xmajorgrids=true,
  extra x ticks={1}, extra x tick labels={},
  extra x tick style={grid=major, major grid style={draw=black!60, dashed}},
  point meta=explicit symbolic,
  bar shift=0pt,
]
\addplot[fill=seriesA, draw=none, nodes near coords, nodes near coords style={font=\tiny, text=black, anchor=east}] coordinates {
  (7.55e-05,1) [7.5e-5]
  (0.000206,2) [2.1e-4]
  (0.000242,3) [2.4e-4]
  (0.00028,4) [2.8e-4]
  (0.000624,5) [6.2e-4]
  (0.00676,6) [0.0068]
  (0.00898,7) [0.0090]
  (0.0102,8) [0.010]
  (0.0111,9) [0.011]
  (0.0116,10) [0.012]
  (0.0289,11) [0.029]
  (0.0355,12) [0.036]
  (0.0383,13) [0.038]
  (0.128,14) [0.13]
  (0.567,15) [0.57]
  (0.639,16) [0.64]
  (0.776,17) [0.78]
  (0.989,18) [0.99]
};
\addplot[fill=seriesA, draw=none, nodes near coords, nodes near coords style={font=\tiny, text=black, anchor=west}] coordinates {
  (1.77,19) [1.8]
  (3.47,20) [3.5]
  (24.2,21) [24]
  (914.6,22) [915]
};
\end{axis}
\end{tikzpicture}
\caption{Application-layer ratio of mean request bytes to mean response bytes, calculated from worksheet 3.22 of the companion workbook~\cite{tr26870}. Values aggregate ten profiles. Provider letters follow the source. The worksheet also contains local-inference runs, labeled separately. Its WebRTC ratios differ from some per-cell ranges in clause C.4.6 of the draft report. The dashed line marks a symmetric exchange.}
\label{fig:uldl}
\end{figure}

The ratios span more than six orders of magnitude. Image generation has a ratio of $7.5\times10^{-5}$. Non-real-time video understanding has a ratio of about 300 in the main local campaign and about 915 in the additional run included in the summary. Input modality and interaction pattern determine directionality. Media uploads can be uplink-heavy. Streamed text and generated images can be downlink-heavy. Basic chat and computer control are closer to balanced. These application ratios describe the recorded serialization. Radio dimensioning also requires protocol overhead and the correct network surface.

The following table shows the per-turn volumes of the personal assistant agent and the agent-to-agent scenarios at the application and network layers.

\begin{table}[H]
\centering
\small
\caption{Per-turn traffic volume of the agentic scenarios at the application layer (driver view) and at the network layer (IP payload in the packet capture).}
\label{tab:agentic-volume}
\begin{tabular}{@{}llrrr@{}}
\toprule
Scenario & Surface and layer & UL (B) & DL (B) & UL/DL \\
\midrule
\multirow{3}{*}{Personal assistant agent} & Control channel, application & 144 & 20,230 & 0.007 \\
 & Control channel, network & 3,817 & 49,981 & 0.076 \\
 & Wide-area egress, network & 175,649 & 353,972 & 0.496 \\
\midrule
Agent-to-agent, all variants & Loopback, network & 1,931 & 4,084 & 0.473 \\
\bottomrule
\end{tabular}
\end{table}

The personal-assistant driver sends a small task to a local runtime and receives a metadata-rich result. This interface has a ratio of 0.007. The runtime's wide-area egress has a ratio of 0.50 and carries about 176~kB of uplink per turn. Its model requests include accumulated context, the system prompt, and tool definitions. The A2A exchanges also carry more bytes at the network layer than their driver payloads suggest. The network counts include transport headers, acknowledgments, and retransmissions. Neither aggregate in the table is uplink-heavy. Capacity estimates need to identify which surface crosses the access network.

The draft report records an initial finding of uplink-heavy agent traffic and notes that repeated context transfer requires further study. Its agentic network measurements give ratios near 0.5. The procedural finding and the measured directionality need to be read with this distinction. Context caching at the inference server can reduce computation~\cite{yuan2026agentic} while the client still sends context bytes. A protocol that retains context on the server may also reduce transmission. The reported volumes characterize the evaluated runtime and client behavior.

\subsection{Burstiness and Delay Sensitivity}

The second assumption states that AI traffic is bursty and that its delay requirement varies within one application. The following table shows the burst descriptors of clause D.4.3.4 of TR 26.870 for the local inference scenarios. Each value is the mean over the eight profiles of the local campaign, which include the \texttt{lossy} profile at 10\% loss, with the observed range in parentheses.

\begin{table}[H]
\centering
\footnotesize
\caption{Burst descriptors of the local inference scenarios, mean over the eight profiles of the local campaign with minimum and maximum in parentheses. The profiles include \texttt{lossy} at 10\% loss.}
\label{tab:bursts}
\setlength{\tabcolsep}{4pt}
\begin{tabularx}{\linewidth}{@{}Xrrr@{}}
\toprule
Scenario & Peak-to-mean ratio & Coefficient of variation & Active duration (ms) \\
\midrule
Video understanding, real-time & 14.35 (13.91 -- 14.82) & 1.62 (1.60 -- 1.64) & 36.1 (34.6 -- 39.7) \\
Video understanding, non-real-time & 29.85 (6.6 -- 82.9) & 2.04 (1.5 -- 2.9) & 8.5 (1 -- 14) \\
Chat, tokens & 11.16 (7.06 -- 17.61) & 1.27 (1.11 -- 1.57) & 65.0 (60.6 -- 70.3) \\
Chat, streaming & 21.18 (9.7 -- 36.7) & 1.18 (1.0 -- 1.4) & 10.9 (8 -- 25) \\
Chat, non-streaming & 38.06 (31 -- 79.4) & 1.75 (1.4 -- 3.5) & 9.5 (9 -- 10) \\
Chat, reasoning & 24.96 (7.2 -- 73.2) & 1.14 (1.0 -- 1.4) & 12.0 (8 -- 33) \\
Multimodal analysis, non-streaming & 13.91 (3.1 -- 79.7) & 1.61 (1.3 -- 3.5) & 18.6 (3 -- 29) \\
\bottomrule
\end{tabularx}
\end{table}

The reported peak-to-mean ratios exceed one on each profile. Several scenarios show substantial variation across profiles, with some maxima under \texttt{lossy}. These ranges measure variation across conditions. They are not confidence intervals or estimates of variation between repeated runs. Real-time video understanding has a smaller relative spread, consistent with a paced uplink. The workbook specifies 1~ms buckets for these descriptors. The active duration is the mean duration of an active burst period. This bucket analysis differs from the gap-based segmentation in Section~\ref{sec:method}. Packet selection also varies by scenario, with token chat excluding small acknowledgment packets. Host scheduling and capture offloads may affect the resulting short-window statistics.

The following table shows how response latency and TTFT change under the \texttt{cell\_edge} and \texttt{congested} profiles relative to the baseline without emulation, for the local inference scenarios.

\begin{table}[H]
\centering
\footnotesize
\caption{Change in response latency under impaired profiles relative to the baseline without emulation, local inference scenarios.}
\label{tab:delay}
\setlength{\tabcolsep}{4pt}
\begin{tabularx}{\linewidth}{@{}Xrrrrr@{}}
\toprule
Scenario & Baseline (s) & Cell edge & Congested & TTFT base (s) & TTFT congested (s) \\
\midrule
Video understanding, non-real-time & 5.02 & +295\% & +889\% & -- & -- \\
Chat, tokens & 6.11 & +1.5\% & +13.7\% & -- & -- \\
Chat, streaming & 6.94 & +15.3\% & +46.4\% & 0.35 & 1.84 \\
Chat, non-streaming & 4.94 & +6.9\% & +24.3\% & -- & -- \\
Chat, reasoning & 15.02 & +16.1\% & +31.3\% & 6.45 & 8.74 \\
Multimodal analysis, streaming & 2.38 & +103.8\% & +216.8\% & 0.23 & 5.32 \\
Multimodal analysis, non-streaming & 9.38 & +17.8\% & +17.3\% & -- & -- \\
\bottomrule
\end{tabularx}
\end{table}

Non-real-time video understanding uploads about 1.3~MB per request. Its response latency rises from 5.02~s to about 49.6~s under \texttt{congested}, whose configured rate is 1~Mbit/s. This profile also changes delay, jitter, and loss. The result does not isolate a bandwidth effect. Smaller requests show different penalties. For streaming multimodal analysis, TTFT grows from 0.23~s to 5.32~s, while total response latency grows from 2.38~s to about 7.54~s. First-output and completion latency respond differently. These measurements motivate reporting both. They do not by themselves define a user delay requirement.

\subsection{Round-Trip Delay and Connection Setup}

The third assumption states that round-trip delay is important. The following table shows the components of round-trip delay aggregated across the remote campaign.

\begin{table}[H]
\centering
\small
\caption{Round-trip delay components aggregated across the remote measurement dataset.}
\label{tab:rtt}
\begin{tabular}{@{}lrrrrr@{}}
\toprule
Component & Observations & p50 (ms) & p95 (ms) & p99 (ms) & Mean (ms) \\
\midrule
TCP handshake RTT & 33,292 & 63.21 & 701.17 & 1,386.49 & 234.45 \\
TLS handshake & 18,604 & 107.68 & 758.21 & 1,947.50 & 272.74 \\
HTTP setup RTT & 32,981 & 0.23 & 38.25 & 313.26 & 24.81 \\
\bottomrule
\end{tabular}
\end{table}

TCP and TLS handshake durations have medians of 63.21~ms and 107.68~ms. Their observation counts differ. Adding their medians or percentiles does not yield the corresponding statistic of end-to-end connection setup. The workbook reports reuse for 4.58\% of TCP flows. This is a flow-level fraction and does not specify the fraction of application requests that required a fresh connection. For the personal assistant, the separate egress summary reports 2.46 flows per turn and 28 distinct destinations across the campaign. Its TLS handshake median is 1.33~s and its 95th percentile is 3.64~s. These observations motivate evaluating persistent connections and TLS resumption. The available aggregates do not quantify their contribution to the critical path of a task.

The clients create fresh sessions between runs. The results emphasize connection establishment and do not represent a continuously reused HTTP/2 or QUIC connection~\cite{rfc9000}. The workbook also reports a median inter-chunk gap of 0.116~ms. Chunk arrival includes server buffering, transport delivery, and client reads. Several events may arrive in one read. Comparing this median with the RTT median cannot isolate model generation time or establish that the network has no effect on token delivery.

\subsection{Variability and Agentic Workloads}

The fourth RAN2 question concerns variability within an application. The workbook reports variation in burst structure, volume, delay, inter-arrival time, reliability, and connection duration. Remote TCP flow duration has a median of 0.71~s and a 95th percentile of 61.5~s. The distinct-destination count per capture has a 95th percentile of 20 and a maximum of 180. These are pooled descriptions across scenarios and profiles. They do not isolate within-application variation at fixed conditions.

The deterministic A2A agents execute without an LLM. Their latency includes protocol handling, transport, configured chunk pacing, and orchestration. The local delegation variant shapes the client-to-orchestrator hop and dispatches downstream tasks sequentially. The following figure depicts mean per-turn latency against twice the nominal one-way delay.

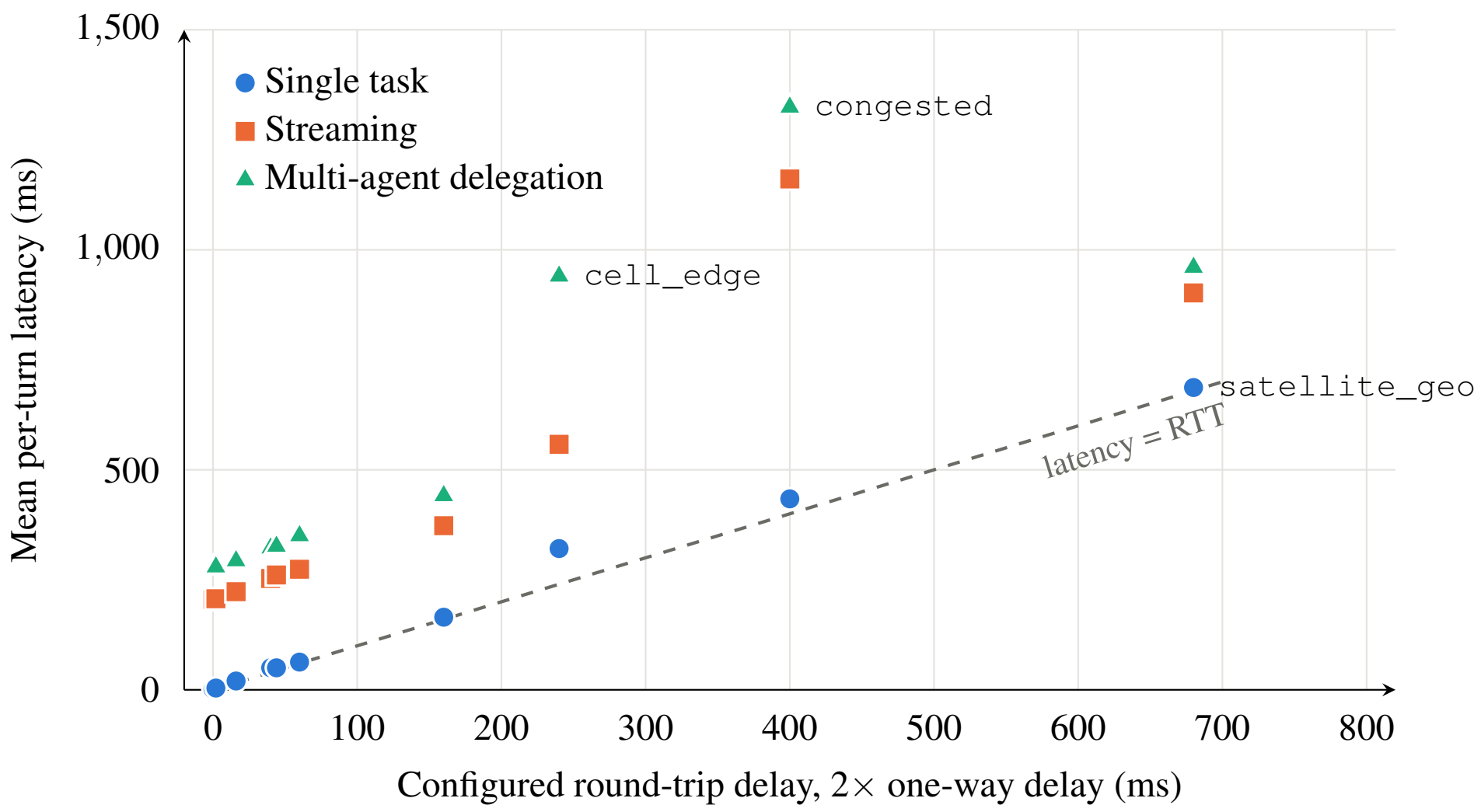
\begin{figure}[!htbp]
\centering
\begin{tikzpicture}
\begin{axis}[
  paperaxis,
  width=0.72\linewidth, height=7.2cm,
  xlabel={Configured round-trip delay, $2\times$ one-way delay (ms)},
  ylabel={Mean per-turn latency (ms)},
  xmin=-20, xmax=820, ymin=0, ymax=1500, clip=false,
  xmajorgrids=true, ymajorgrids=true,
  legend pos=north west, legend cell align=left,
  every axis plot/.append style={mark size=2.6pt},
]
\addplot[draw=inkmuted, dashed, line width=0.8pt, mark=none, domain=0:700, forget plot] {x};
\node[font=\scriptsize, text=inkmuted, rotate=17] at (axis cs:640,560) {latency = RTT};
\addplot[only marks, mark=*, color=seriesA, mark options={draw=white, line width=0.6pt}] coordinates {
  (0,2) (2,4) (40,50) (60,63) (240,321) (44,50) (680,687) (400,434) (160,165) (16,20)
};
\addlegendentry{Single task}
\addplot[only marks, mark=square*, color=seriesB, mark options={draw=white, line width=0.6pt}] coordinates {
  (0,205) (2,207) (40,253) (60,274) (240,558) (44,261) (680,902) (400,1161) (160,373) (16,223)
};
\addlegendentry{Streaming}
\addplot[only marks, mark=triangle*, color=seriesC, mark size=3.2pt, mark options={draw=white, line width=0.6pt}] coordinates {
  (0,277) (2,279) (40,322) (60,350) (240,940) (44,326) (680,960) (400,1324) (160,441) (16,293)
};
\addlegendentry{Multi-agent delegation}
\node[font=\scriptsize, anchor=west] at (axis cs:686,687) {\texttt{satellite\_geo}};
\node[font=\scriptsize, anchor=west] at (axis cs:406,1324) {\texttt{congested}};
\node[font=\scriptsize, anchor=west] at (axis cs:246,940) {\texttt{cell\_edge}};
\end{axis}
\end{tikzpicture}
\caption{Mean per-turn latency of the agent-to-agent scenarios versus the configured round-trip delay of each network profile. The deterministic agents execute without model inference. Streaming includes configured chunk spacing. Delegation shapes only the client-to-orchestrator hop. Data from draft TR 26.870~\cite{tr26870}.}
\label{fig:a2a}
\end{figure}

Single-task latency follows the nominal round-trip delay over much of the profile range. It reaches 687~ms on \texttt{satellite\_geo}, compared with a nominal 680~ms. Streaming adds chunk spacing. Delegation has a baseline of 277~ms that includes orchestration. The largest upward deviations occur on \texttt{cell\_edge} and \texttt{congested}. These profiles combine several impairments, and the experiment does not separate their causes. Failures occur only on these two profiles. This result motivates further study of delay and reliability for agent signaling. It does not establish a packet-error-rate target or a common QoS requirement for agent systems.

The personal assistant has a mean task latency of 11.4~s without added impairment and 9.7~s on \texttt{6g\_itu\_hrllc}. The unshaped tasks range from 5.4~s to 31.2~s. This spread prevents attributing the lower mean to the latter profile. Mean latency rises to 24.4~s on \texttt{satellite\_geo} and 26.6~s on \texttt{congested}. All 200 reported turns complete successfully. Long timeouts and reliable transport are consistent with completion despite impairment. The observed success rate does not establish a general reliability guarantee.

\subsection{Tokenized Traffic}

The fifth question concerns tokenized traffic. A model token is a unit of the model representation. An application chunk is a unit delivered by an interface and may contain several tokens. A network packet is a transport unit and may contain all or part of a chunk. Streamed text exposes incremental output at the application interface. The evaluated video client instead tokenizes sampled frames and sends discrete identifiers over RTP. Responses return through a WebRTC data channel. The following table shows token metrics for the initial video campaign and the token-identifier chat scenario without added impairment.

\begin{table}[H]
\centering
\small
\caption{Token-related metrics for the scenarios with tokenized traffic, profile without emulation.}
\label{tab:tokens}
\begin{tabular}{@{}lll@{}}
\toprule
Metric & Video understanding, real-time & Chat, tokens \\
\midrule
Uplink token rate (token/s) & 2,051 & Not resolved in source \\
Downlink output token rate (token/s) & 24.1 & 214 \\
Uplink encapsulation & RTP packets & HTTP token array \\
Downlink encapsulation & Streamed chunks & HTTP response body \\
Uplink volume per token (B) & about 2.5 & about 6.0 \\
Burst peak-to-mean ratio & 13.91 & 16.11 \\
Active / idle duration (ms) & 36.7 / 2,578 & 70.3 / 1,258 \\
\bottomrule
\end{tabular}
\end{table}

The combined workbook summary labels about 200~token/s as the chat uplink rate, but the scenario sheet identifies 200 to 218~token/s as output rates. We leave the uplink rate unreported. The two video campaigns report about 2.2 to 2.5~B per generated uplink token without added impairment. The accounting includes packetization overhead and is consistent with compact token identifiers. A two-byte identifier can address 65,536 vocabulary entries. An embedding instead carries a vector whose byte count depends on dimension and precision. The term tokenized traffic needs to identify the representation. The chat fits report 4.4 to 5.4~request bytes per input token with $r^2 \geq 0.996$. Streamed response fits report 398 to 440~B per output token, including event framing and metadata in these implementations. These slopes do not define the storage size of a model token.

Huawei contributed an extended campaign~\cite{s4261818} using Liquid~\cite{liquid} and VILA-U~\cite{vilau} on 25 StreamingBench clips~\cite{streamingbench}, with five questions per clip. The workbook reports 125 questions per profile and model. This campaign uses different model and response configurations from the initial campaign. The following figure depicts its delivered uplink volume and task accuracy under independent uplink loss.

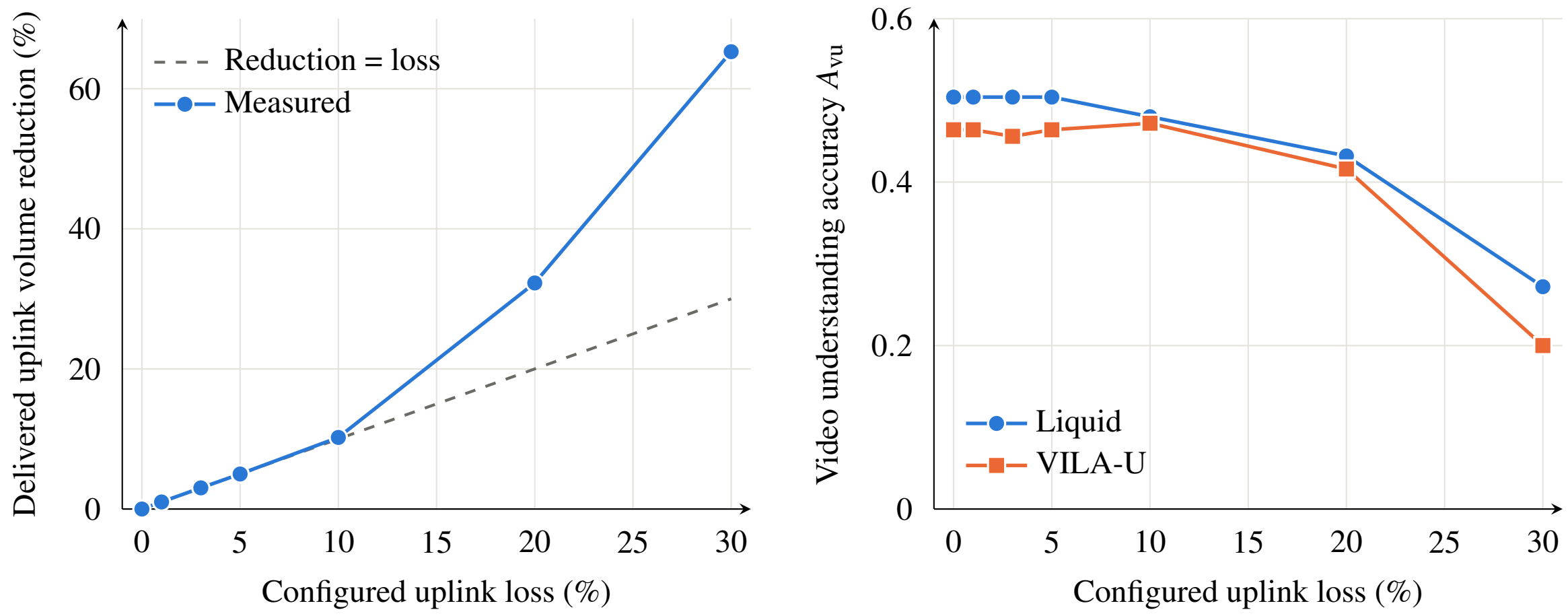
\begin{figure}[!htbp]
\centering
\begin{tikzpicture}
\begin{groupplot}[
  group style={group size=2 by 1, horizontal sep=18mm},
  paperaxis,
  width=0.47\linewidth, height=6.4cm,
  xmin=-1, xmax=31,
  xlabel={Configured uplink loss (\%)},
  xmajorgrids=true, ymajorgrids=true,
  every axis plot/.append style={mark size=2.4pt},
]
\nextgroupplot[
  ylabel={Delivered uplink volume reduction (\%)},
  ymin=0, ymax=70,
  legend pos=north west, legend cell align=left,
]
\addplot[draw=inkmuted, dashed, line width=0.8pt, mark=none, domain=0:30] {x};
\addlegendentry{Reduction = loss}
\addplot[color=seriesA, mark=*, mark options={draw=white, line width=0.6pt}] coordinates {
  (0,0) (1,1.01) (3,3.02) (5,5.00) (10,10.22) (20,32.27) (30,65.28)
};
\addlegendentry{Measured}
\nextgroupplot[
  ylabel={Video understanding accuracy $A_{\mathrm{vu}}$},
  ymin=0, ymax=0.6,
  legend pos=south west, legend cell align=left,
]
\addplot[color=seriesA, mark=*, mark options={draw=white, line width=0.6pt}] coordinates {
  (0,0.504) (1,0.504) (3,0.504) (5,0.504) (10,0.480) (20,0.432) (30,0.272)
};
\addlegendentry{Liquid}
\addplot[color=seriesB, mark=square*, mark options={draw=white, line width=0.6pt}] coordinates {
  (0,0.464) (1,0.464) (3,0.456) (5,0.464) (10,0.472) (20,0.416) (30,0.200)
};
\addlegendentry{VILA-U}
\end{groupplot}
\end{tikzpicture}
\caption{Real-time video understanding under independent uplink packet loss. Left: reduction of delivered uplink volume relative to the baseline. Right: correct answers divided by 125 questions per model and profile. Lines connect the reported observations and do not represent a fitted model. Repeated-run uncertainty is unavailable. Data from Annex D of TR 26.870~\cite{tr26870}, contributed in~\cite{s4261818}.}
\label{fig:loss}
\end{figure}

Through 10\% configured loss, uplink volume reduction is close to the configured fraction. It reaches 32.27\% at 20\% loss and 65.28\% at 30\% loss. The generated token count also falls at the two highest settings. The volume reduction includes session behavior in addition to packet dropping. Through 10\% loss, each model changes by at most three correct answers relative to baseline. At 20\%, Liquid falls from 63 to 54 correct answers, or from 50.4\% to 43.2\%. VILA-U falls from 58 to 52, or from 46.4\% to 41.6\%. These are declines of 7.2 and 4.8 percentage points. At 30\%, accuracy is 27.2\% and 20.0\%, near the 25\% chance reference for four choices. The workbook does not provide repeated-run dispersion or paired question outcomes. Questions from the same clip may also be dependent. The observations do not establish equivalence at 20\% or a precise loss-tolerance boundary.

The workbook reports about 3.27 output tokens per application chunk across the tested profiles. This relation may help estimate token counts at an instrumented application interface. Encrypted network packets do not expose these chunk boundaries directly. The relation also depends on the response implementation. Correlated-loss tolerance remains unmeasured. Its effect can depend on packetization, temporal dependencies, recovery, and task sampling, so independent loss does not provide a general bound.

\subsection{Synthetic Traffic Generation}

The retained dataset snapshot contains 56,297 attributed flows in 927 capture groups, derived from 1,960 captures and 8,022 labeled sessions~\cite{testbedrepo}. It includes 19,729 loopback flows and 164 User Datagram Protocol (UDP) flows. Streaming contributes 16,259 flows, while the real-time WebRTC category contains nine. These counts describe the dataset snapshot, rather than the remote campaign alone. Two category-conditioned Markov baselines use the clean-profile training partition. Order 0 samples channel marginals independently. Order 1 adds transitions within the size, inter-arrival, and direction channels, while retaining independence between them. Both decode bins by sampling the empirical training values within each bin. The following table shows the two-sample Kolmogorov-Smirnov (KS) statistic~\cite{massey1951ks} against 931 held-out clean-profile flows in seven categories, as reported with the modeling study~\cite{s4261447}.

\begin{table}[H]
\centering
\footnotesize
\caption{Held-out KS statistic of the Markov baselines, unweighted mean over seven traffic categories. Lower is better.}
\label{tab:ks}
\begin{tabularx}{\linewidth}{@{}Xrrrrrr@{}}
\toprule
Generator & Packet size & Inter-arrival & Sequence length & Uplink fraction & Total bytes & Duration \\
\midrule
Markov order 0 & 0.134 & 0.142 & 0.331 & 0.286 & 0.513 & 0.387 \\
Markov order 1 & 0.131 & 0.136 & 0.331 & 0.284 & 0.477 & 0.373 \\
\bottomrule
\end{tabularx}
\end{table}

Packet-size and inter-arrival KS distances are about 0.13 to 0.14. Flow-level distances are larger, reaching 0.51 for total bytes. Separate channel chains do not capture the joint relationship between timing, size, and direction. Sequence length is sampled from the empirical training distribution. Its held-out discrepancy can include partition differences, finite sampling, and the sequence-length cap. These distances compare empirical distributions. They are not significance tests over independent packets. Capture-group separation reduces direct leakage but leaves temporal and provider generalization for further evaluation. A learned generator needs comparison against these baselines on the same partitions and accounting rules.
\section{Emulator Fidelity and Threats to Validity}
\label{sec:fidelity}

A profile identifies emulator inputs. The resulting path statistics depend on the queueing arrangement, workload, and implementation revision. ATEME examined the common platform described in TR 26.934~\cite{tr26934} during QUIC media-delivery tests~\cite{s4ai260095}. We compare that analysis with the study configuration and the retained emulator sources. The ATEME snapshot differs in distribution, explicit reordering, and some loss parameters. The results below identify these differences rather than treating the catalog as one invariant configuration.

\subsection{Reordering under Per-Packet Jitter}

Netem assigns per-packet delivery times from a delay distribution. Packets can overtake one another when their spacing is smaller than the delay variation. A delay-correlation setting does not provide an in-order service guarantee. For the table-based distributions in the checked Linux 6.1 and 6.17 kernels, reduction of the random value to a distribution-table index can attenuate the requested correlation. The retained probes report lag-1 delay autocorrelation within $\pm 0.03$ for settings of 0\%, 50\%, and 99\%. This result is specific to those versions and probes. The following table shows ATEME's reordering measurements for bursts of eight 1,200-byte packets every 40~ms, equivalent to 1.92~Mbit/s of packet data.

\begin{table}[H]
\centering
\footnotesize
\caption{Reordering through the emulator with HTB rate limiting and per-packet netem jitter, measured in~\cite{s4ai260095} on the earlier profile snapshot. Delay and jitter magnitudes match \cref{tab:profiles}, but the earlier cell-edge distribution was Pareto rather than Paretonormal and its explicit reordering setting differed.}
\label{tab:reorder}
\begin{tabularx}{\linewidth}{@{}Xrrr@{}}
\toprule
Profile & Median one-way delay (ms) & Arrived out of order & Overtaken by a packet $\geq 3$ later \\
\midrule
\texttt{cell\_edge} & 110.5 & 52.7\% & 33.9\% \\
\texttt{wifi\_good} & 30.0 & 65.7\% & 44.3\% \\
\texttt{5g\_urban} & 20.1 & 64.7\% & 42.7\% \\
\texttt{5qi\_80} & 8.0 & 64.5\% & 42.1\% \\
\bottomrule
\end{tabularx}
\end{table}

The Packet Data Convergence Protocol (PDCP) supports in-sequence delivery, subject to configuration and reordering-timer behavior~\cite{ts38323}. An emulator needs to state whether it represents an ordered service. RFC 9002~\cite{rfc9002} recommends an initial QUIC packet threshold of three and permits adaptation. Receiver arrival order alone does not determine sender loss declarations. ATEME's tested QUIC relay retransmitted 22\% to 26\% of packets on its \texttt{cell\_edge} configuration, with 77\% to 84\% of those retransmissions classified as spurious. This result concerns that implementation and configuration. TCP implementations using Recent ACKnowledgment and Tail Loss Probe (RACK-TLP)~\cite{rfc8985} can adapt to reordering, but the resulting effect on the AI campaigns was not measured. Netem's native rate scheduling can serialize queued packets and reduce reordering. It also changes delay. ATEME reports a cell-edge median rising from about 110~ms to 167~ms. The loopback path already uses native rate scheduling when a rate is configured, so the main-interface reordering percentages do not transfer directly to local experiments.

\subsection{Suppression of Correlated Loss}

Netem's correlated random-loss generator mixes a fresh random value with its previous state. For small nominal loss probabilities, the resulting distribution can produce far fewer threshold crossings. We evaluate the integer recurrence offline using $10^7$ draws per setting from a deterministic SplitMix64 input stream. The ancillary files provide the implementation, its fixed seeds, and the resulting event counts. This is an algorithm check and does not transmit packets through the kernel. The following table shows selected results.

\begin{table}[H]
\centering
\small
\caption{Loss events of the netem correlated loss generator in $10^7$ draws, offline evaluation of the kernel recurrence.}
\label{tab:loss-suppression}
\begin{tabular}{@{}lrr@{}}
\toprule
Setting & Loss events & Effective rate \\
\midrule
Loss 0.1\%, correlation 25\% (\texttt{5g\_urban}) & 0 & 0 observed \\
Loss 0.1\%, correlation 30\% (\texttt{wifi\_good}) & 0 & 0 observed \\
Loss 0.1\%, correlation 0\% (reference) & 10,074 & 0.10074\% \\
Loss 5\%, correlation 15\% & 71,012 & 0.71012\% \\
\bottomrule
\end{tabular}
\end{table}

The checked correlated-loss inputs produce no loss events in $10^7$ draws for \texttt{6g\_itu\_hrllc}, \texttt{5g\_urban}, \texttt{wifi\_good}, both satellite directions, \texttt{5qi\_7}, and \texttt{5qi\_80}. ATEME also reports no loss in 50,000 packets at 0.1\% nominal loss and 15\% correlation. With zero correlation, it reports 51 losses in 49,950 packets~\cite{s4ai260095}. These observations do not establish zero loss probability. Queue overflow and the external path can add loss. They show that a configured nominal rate is not a measured path rate. Independent loss can be requested by setting correlation to zero in the correlated-random-loss syntax. This change must not be applied to Gilbert-Elliott syntax, where the second parameter has a different meaning.

\subsection{Gilbert-Elliott Parameter Mapping}

The \texttt{cell\_edge} and \texttt{congested} profiles use a Gilbert-Elliott model~\cite{gilbert1960,elliott1963}. With loss probability 1 in the bad state and 0 in the good state, the model is defined by the transition probability $p$ from the good to the bad state and $r$ from the bad to the good state. The stationary loss rate $L$ and the mean loss burst length $B$ are
\begin{equation}
  L = \frac{p}{p + r}, \qquad B = \frac{1}{r}.
\end{equation}
The inspected emulator revision maps the configured loss rate to $p$ and the field named loss correlation to $r$. For \texttt{cell\_edge}, $p=1\%$ and $r=35\%$ imply a stationary loss rate of 2.78\%. For \texttt{congested}, $p=3\%$ and $r=40\%$ imply 6.98\%. These are model expectations, not measured path losses. Earlier snapshots supplied only the first parameter, resulting in an independent-loss configuration. ATEME reports 0.8\% to 1.2\% on that earlier cell-edge case. The available campaign provenance does not identify the mapping used by every workbook result. Setting the second parameter to zero would make the bad state absorbing and is not a correction. For a target stationary loss rate $L$ and mean burst length $B$, the parameters are
\begin{equation}
  r = \frac{1}{B}, \qquad p = \frac{L\, r}{1 - L}.
\end{equation}
Preserving the burst lengths implied by the current settings, this gives $p = 0.3535\%$ for \texttt{cell\_edge} and $p = 1.2371\%$ for \texttt{congested}.

\subsection{Consequences for the Published Results}

The independent-loss sweep does not use the correlated-random-loss or Gilbert-Elliott mappings. This removes those two specific ambiguities, but it does not verify receiver loss or eliminate queue effects. Deployment-profile results require the executed configuration and queueing path for interpretation. Cross-scenario volume asymmetry remains an observation of the recorded campaigns. Its magnitude under corrected profiles requires new measurements. A global severity ordering cannot be inferred when profiles change delay, rate, ordering, and loss together. Quantitative claims about configured loss need repetition with measured loss and declared byte accounting.

\subsection{Other Threats to Validity}

The following threats apply in addition to the emulator effects:
\begin{itemize}
  \item \textbf{Provider-side variance.} For remote execution, the variance of the provider can exceed the effect of the network. For chat with provider A, the mean latency without emulation is 19.5~s and on \texttt{5g\_urban} 9.8~s. The orchestrator runs all runs of one profile before the next, so a drift in provider load over the hours of a campaign is confounded with the profile. Interleaving profiles in randomized order would separate the two.
  \item \textbf{Application-layer volumes.} Request and response sizes in the session records are estimated from the serialized payloads. They exclude transport and TLS overhead and, for agents, the traffic of the runtime itself. Statements on directionality for radio dimensioning need the network-layer volumes.
  \item \textbf{Transport coverage.} Most hosted requests use TCP. WebRTC media use UDP. The campaign does not provide a matched comparison with QUIC or persistent connections. Filters and host segmentation or aggregation offloads can also change the packet units visible in a capture.
  \item \textbf{Evaluation support.} Tokenized video uses one transport implementation with two models and 125 questions per profile. Repeated-run dispersion and paired outcomes are unavailable. The agentic measurements cover one runtime and deterministic A2A agents.
  \item \textbf{Aggregation across profiles.} Some local aggregates include the \texttt{lossy} stress case. Cross-profile ranges do not estimate uncertainty at a fixed condition. Retry and resume rules can also change the population of completed records. Reliability estimates require all attempts and declared exclusion rules.
\end{itemize}

The profile contract needs to distinguish intrinsic generator parameters from observed path statistics. It also needs to state the queueing path, ordering behavior, loss process, byte accounting, and workload. Qualification should report delay distributions, loss and burst lengths, rate, reordering, and overflow under declared packet sizes and pacing. Corrected campaigns need a new catalog version. Earlier labels and raw records need to remain associated with the configuration that produced them.
\section{Use in Standardization}
\label{sec:standards}

The testbed was introduced to SA4 in February 2026~\cite{s4260114}. It is documented as Annex B of TR 26.870 as the testbed used to obtain the measurements of the study, with the network emulator documented in Annex E. The network profiles were agreed in S4-260848~\cite{s4260848}, and the methodology for the RAN2 questions in S4-260859~\cite{s4260859}. The hosting of the software as a 5G-MAG reference tool was addressed in S4-261323~\cite{s4261323}. The source code is available under the 5G-MAG public license~\cite{testbedrepo}.

Participating companies extended scenarios and reviewed changes through the public repository. Huawei contributed real-time video understanding and token-identifier scenarios. The draft report records five initial findings in clause 7.1 with references to Annex C. These findings remain part of an evolving study. Our analysis identifies where pooled measurements, observation points, and emulator revisions limit their interpretation. Qualcomm also contributed high-level observations on scenario dependence and token representation~\cite{s4ap260111}. Kundu et al.~\cite{kundu2026ainative} use the testbed in a separate discussion of AI traffic awareness.

\subsection{Software and Data Availability}

The testbed source is available from 5G-MAG~\cite{testbedrepo}. The implementation review uses the refactor revision \texttt{d055fa68ef4d}. The report and workbook snapshot is S4aP260148, distributed on October 23, 2026~\cite{tr26870}. The figures in this paper use values from that workbook. The source code and nominal profiles alone do not reproduce historical hosted-service behavior. Reproduction of a campaign also requires its model configuration, prompts, executed commands, software versions, captures, and session records. This paper does not distribute the raw hosted-service captures. The dataset counts and generator statistics describe a retained study snapshot and do not constitute a released benchmark.
\section{Summary}
\label{sec:conclusion}

We described an open testbed for measuring AI service traffic and analyzed the results recorded in draft TR 26.870 V0.6.2. The testbed separates network emulation from scenario execution and associates application records with packet captures. The reported traffic has different directional ratios, burst structures, and delay behavior across scenarios. Agent measurements show that local driver traffic and wide-area runtime traffic can have different volumes. The deterministic A2A experiments show the effect of path delay without model inference. The tokenized-video sweep shows accuracy degradation at high independent loss, with limited support for estimating a tolerance boundary.

The emulator analysis identifies three interpretation issues: reordering under per-packet jitter, suppression of rare correlated-random-loss events, and a Gilbert-Elliott parameter mapping that changes stationary loss. Main-interface and loopback shaping also differ. A nominal profile does not establish the conditions experienced by an application. Campaigns need versioned configurations and workload-specific path measurements.

Further work includes repeated campaigns with corrected profiles and randomized profile order, matched evaluation of persistent TCP and QUIC connections, and experiments with correlated or trace-driven loss. Task-quality evaluation needs repeated runs and per-question outcomes. Synthetic traffic generation needs comparison on joint flow statistics and held-out campaigns. These steps would allow the measurements to support more precise traffic models and network requirements.

\section*{Acknowledgments}

The author thanks the 5G-MAG reference tools team for hosting, project management and code review of the testbed, Huawei for the real-time video understanding scenario and its loss sweep, ATEME for the analysis of the netem configuration, and the delegates of 3GPP SA4 who contributed, reproduced and reviewed the results collected in TR 26.870.

\bibliographystyle{unsrtnat}
\bibliography{refs}

\end{document}